\documentclass[5p,twocolumn,times,number]{elsarticle}

\usepackage{lineno}

\usepackage{graphicx}
\usepackage{amsmath}

\begin{document}

\begin{frontmatter}

\title{Modeling of Radiation Damage Effects in Silicon Detectors at High Fluences HL-LHC with Sentaurus TCAD}

\author[add1,add2]{D.~Passeri\corref{cor}}
\ead{daniele.passeri@unipg.it}
\author[add3,add2]{F. Moscatelli}
\author[add1,add2]{A. Morozzi}
\author[add2]{G.M. Bilei}

\cortext[cor]{Corresponding author}

\address[add1]{Universit\`a degli Studi di Perugia, Dipartimento di Ingegneria, Perugia, Italy}
\address[add2]{Istituto Nazionale di Fisica Nucleare, Sezione di Perugia, Italy}
\address[add3]{Istituto per la Microelettronica e i Microsistemi, CNR Bologna, Italy}

\begin{abstract}
In this work we propose the application of an enhanced radiation damage model based on the introduction of deep level traps / recombination centers suitable for device level numerical simulation of silicon detectors at very high fluences (e.g. 2.0 $\times$ 10$^{16}$ 1 MeV equivalent neutrons/cm$^2$). We present the comparison between simulation results and experimental data for $p$-type substrate structures in different operating conditions (temperature and biasing voltages) for fluences up to 2.2 $\times$ $10^{16}$  neutrons/cm$^{2}$. The good agreement between simulation findings  and experimental measurements fosters the application of this modeling scheme to the optimization of the next silicon detectors to be used at HL-LHC.

\end{abstract}


\end{frontmatter}

\section{Introduction}

In the last few decades, extensive experimental and  simulation studies have 
been carried out to understand the mechanisms of radiation damage in silicon sensors to be used in High-Energy Physics experiments. In particular, we developed and validated a TCAD model based on the introduction of three deep-level traps able to reproduce the radiation damage macroscopic effects up to fluences in the order of 1 $\times$ $10^{15}$  neutrons/cm$^{2}$ (\cite{passeri:2001}, \cite{petasecca:2006}). However, the new fluences expected at the HL-LHC impose new challenges and the extension of the model is not straightforward. New effects have to be taken into account (e.g. avalanche multiplications and capture cross section dependencies on temperature and fluences), at the same time keeping the solid physically based approach (e.g. by using no extra fitting parameters). In this work we propose the application of an enhanced radiation damage model still based on the introduction of deep level traps / recombination centers suitable for device level numerical simulation of silicon detectors at very high fluences (e.g. 2.2 $\times$ $10^{16}$  neutrons/cm$^{2}$).

\section{Simulation model and results}

\begin{table}
\begin{tabular}{|l|l|l|l|c|}
\hline
\rule[-2mm]{0mm}{6mm}
Defect & E (eV)  & $\sigma_{e}$ (cm$^{-2})$ &  $\sigma_{n}$ (cm$^{-2})$ & $\eta $\\
\hline
\hline
\rule[-1.5mm]{0mm}{5.5mm}
Acceptor  &   $E_{c} - 0.42$   &  1.00$\times 10^{-15}$  &  1.00$\times 10^{-14}$ & 1.6\\
\hline
\rule[-1mm]{0mm}{5.5mm}
Acceptor  &   $E_{c} - 0.46$   &  7.00$\times 10^{-15}$  &  7.00$\times 10^{-14}$ & 0.9\\
\hline
\rule[-1mm]{0mm}{5.5mm}
Donor  &   $E_{v} + 0.36$   &  3.23$\times 10^{-13}$  &  3.23$\times 10^{-14}$ & 0.9\\
\hline
\end{tabular}
\caption{Parameters for fluences up to 7 $\times 10^{15}$ n/cm$^{2}$.}
\label{tab:parI}
\end{table}

\begin{table}
\begin{tabular}{|l|l|l|l|c|}
\hline
\rule[-2mm]{0mm}{6mm}
Defect & E (eV)  & $\sigma_{e}$ (cm$^{-2})$ &  $\sigma_{n}$ (cm$^{-2})$ & $\eta $\\
\hline
\hline
\rule[-1mm]{0mm}{5.5mm}
Acceptor  &   $E_{c} - 0.42$   &  1.00$\times 10^{-15}$  &  1.00$\times 10^{-14}$ & 1.6\\
\hline
\rule[-1mm]{0mm}{5.5mm}
Acceptor  &   $E_{c} - 0.46$   &  3.00$\times 10^{-15}$  &  3.00$\times 10^{-14}$ & 0.9\\
\hline
\rule[-1mm]{0mm}{5.5mm}
Donor  &   $E_{v} + 0.36$   &  3.23$\times 10^{-13}$  &  3.23$\times 10^{-14}$ & 0.9\\
\hline
\end{tabular}
\caption{Parameters for fluences within 7 $\times 10^{15}$ n/cm$^{2}$ and 2.2 $\times 10^{16}$ n/cm$^{2}$.}
\label{tab:parII}
\end{table}

A comprehensive analysis of the variation of the 
charge collection efficiency as a function of the fluence of a sample $n$-in-$p$ strip has been performed using the Synopsys Sentaurus TCAD device simulator. The analysis is based on a past modeling scheme featuring three levels and slightly increased introduction rate for the acceptor level closest to mid-gap to cope with direct inter-defect charge exchange, which was successfully adopted for the optimization of the silicon detectors operating at LHC \cite{petasecca:2006}. In order to extend the suitability of the model, an additional e/h  pair generation due to avalanche effect (impact ionization) has been considered, while the bulk radiation induced defects have been parametrized as reported in Tables \ref{tab:parI} and \ref{tab:parII}.

\begin{figure}
\vskip-5mm
\includegraphics[width=0.95\linewidth]{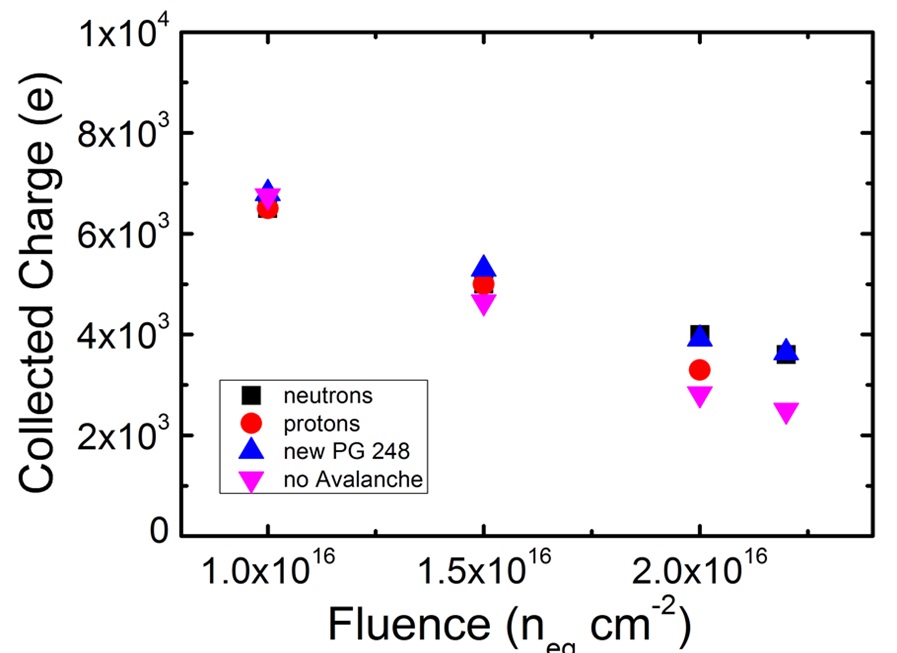}
\vskip-2mm
\caption{Charge collection at T=248K, V$_{BIAS}$=900V: comparison between simulations and measurements (data taken from \cite{lozano:2005}, \cite{affolder:2010}) without avalanche multiplication effects.}
\label{fig:cce_no_av}
\end{figure}

\begin{figure}
\hskip10mm
\vskip-5mm
\includegraphics[width=0.99\linewidth]{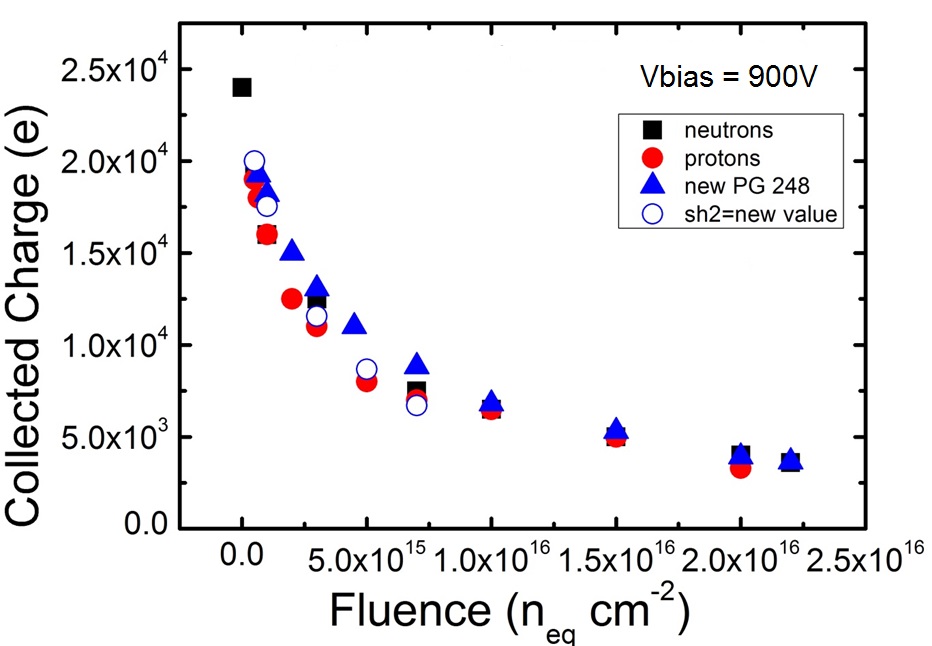}
\vskip-2mm
\caption{Charge collection at T=248K, V$_{BIAS}$=900V considering the impact ionization and enhanced capture cross-sections.}
\label{fig:cce_av}
\end{figure}

The simulated active behavior of irradiated detectors has been 
compared with experimental measurements extracted from the literature (\cite{lozano:2005}, \cite{affolder:2010}), showing a very good agreement when
the proper models of avalanche generation and the variations of capture cross sections as a funcion of the irradiation fluence are taken into account. In particular in Figure \ref{fig:cce_no_av}  the comparison between simulated and experimental charge collection of a sample $n$-in-$p$ strip detectors at 248 K is reported. If the avalanche generation is not taken into account, a marked underestimation of the collected charge is obtained at high fluences. 
On the other hand, a very good agreement along all the full expected range of operation at HL-LHC has been obtained by considering the avalanche generation (in particular the Van Overstraeten model) and the near mid-gap acceptor cross section variation as illustrated in Figure \ref{fig:cce_av}.

\begin{figure}
\centering
\vskip-3mm
\includegraphics[width=0.70\linewidth]{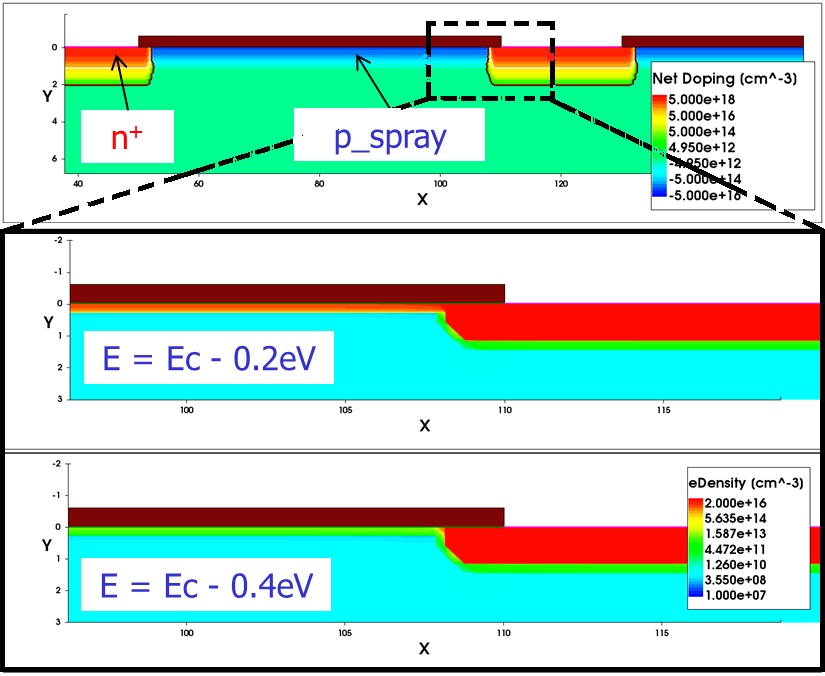}
\vskip2.5mm
\caption{Net doping concentrations and inversion electron layer as a function of the interface acceptor trap state ($\Phi$=2.0 $\times 10^{16}$ n/cm$^{2}$) .}
\label{fig:2strip}
\end{figure}

\begin{figure}
\hskip10mm
\vskip-2.5mm
\includegraphics[width=0.95\linewidth]{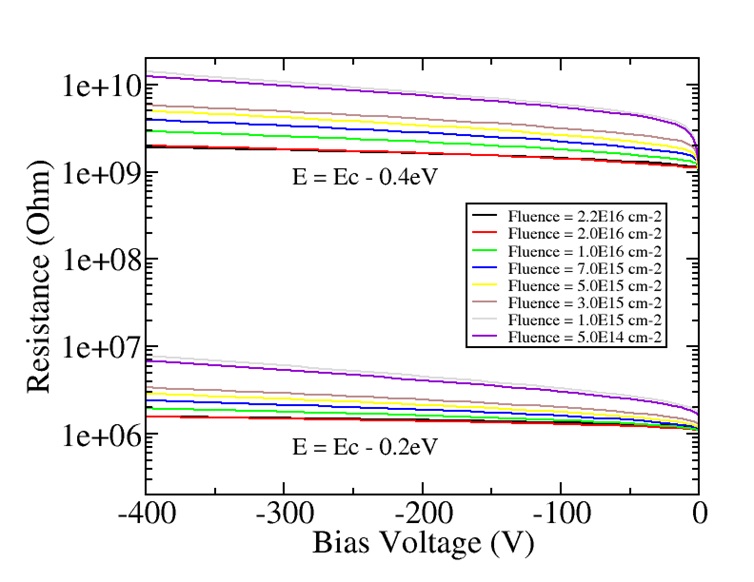}
\caption{Interstrip resistance as a function of the V$_{BIAS}$ at different fluences and for different energies of the interface trap level.}
\label{fig:Rint}
\end{figure}

In order to develop a comprehensive bulk and surface damage model, the fixed oxide-charge density increase and additional interface trap state build-up with irradiation fluence have been considered as well. To evaluate the surface effect in term of strip isolation, a simple two strip structure with relatively low doped $p$-spray isolation layer has ben considered (Figure \ref{fig:2strip}). 
In particular, the effect of energy level of a single Si/SiO$_{2}$ interface  acceptor trap state is illustrated in Figure \ref{fig:2strip}: the electron conduction (inversion layer) path suppression increases the interstrip resistance. A more comprehensive picture of the effect is illustrated in Figure \ref{fig:Rint}, where 
the interstrip resistance as a function of the fluence for different position of the interface acceptor trap state is reported. 

\section{Conclusion}
A new damage modelling scheme, suitable within commercial TCAD tools, has been proposed, aiming to extend its application to high fluences HL-LHC radiation damage levels (e.g. fluences  $>$ 2.2 $\times 10^{16}$ n/cm$^{2}$). This model should be therefore used as a predictive tool for investigating sensor behavior at different fluences, temperatures, and bias voltages and will be used for the optimization of 3D and planar silicon detectors for future HL-LHC High Energy Physics experiments.

\begin{thebibliography}{10}
\expandafter\ifx\csname url\endcsname\relax
  \def\url#1{\texttt{#1}}\fi
\expandafter\ifx\csname urlprefix\endcsname\relax\def\urlprefix{URL }\fi

\bibitem{passeri:2001}
D.~Passeri, P.~Ciampolini, G.M.~Bilei, and F.~Moscatelli, Comprehensive Modeling of Bulk-Damage Effects in Silicon Radiation Detectors, IEEE Trans. on Nuclear Science, vol. 48, no. 5, October 2001.

\bibitem{petasecca:2006}
M.~Petasecca, F.~Moscatelli, D.~Passeri, and G.~U.~Pignatel, Numerical Simulation of Radiation Damage Effects in p-Type and n-Type FZ Silicon Detectors, IEEE Trans. on Nuclear Science, vol. 53, no. 5, October 2006.

\bibitem{lozano:2005}
M.~Lozano, et~al., Comparison of radiation hardness of P-in-N, N-in-N, and N-in-P silicon pad detectors, IEEE Trans. Nucl. Sci. 52 (5) (2005).

\bibitem{affolder:2010}
A.~Affolder, et~al., Collected charge of planar silicon detectors after pion and proton irradiations up to 2.2x1016 n/cm2, NIM A, Vol. 623 (2010).


\end{thebibliography}
\end{document}